\documentclass[a4paper]{jpconf}
\usepackage{graphicx}
\usepackage{amssymb}
\usepackage{float}
\usepackage{array}
\usepackage{xcolor} 
\usepackage{ulem}

\begin{document}
\title[Towards a sub-kelvin cryogenic Fabry-Perot silicon cavity]
{Towards a sub-kelvin cryogenic Fabry-Perot silicon cavity}
\author{Joannès Barbarat$^{1}$, Jonathan Gillot$^{2}$, Jacques Millo$^{2}$, Clément Lacroûte$^{1}$, Thomas Legero$^{3}$, Vincent Giordano$^{1}$ and Yann Kersalé$^{2}$ }

\address{$^{1}$Université de Franche-Comté, CNRS, FEMTO-ST, 26 rue de l’Epitaphe, 25000 Besançon, France\\
$^{2}$ENSMM, CNRS, FEMTO-ST, 26 rue de l’Epitaphe, 25000 Besançon, France\\
$^{3}$Physikalisch-Technische Bundesanstalt, Bundesallee 100, 38116 Braunschweig, Germany}
\ead{jonathan.gillot@femto-st.fr}
\vspace{10pt}

\begin{abstract}
We report on the development of a sub-kelvin, single-crystal silicon Fabry-Perot cavity. Operating such a cavity below 1~K should reduce the thermal noise limit of the cavity, and by this way address the current limitations of ultrastable lasers. To further decrease mechanical losses, mirrors with silicon substrates and crystalline coatings are optically contacted to the spacer, resulting in a room-temperature finesse of 220,000. To operate our cavity at sub-kelvin temperatures, we use a dilution refrigerator able to reach temperatures down to 10 mK. We have designed a mechanical mount to house our cavity in such a cryostat, with optimized heat transfer that will decrease the cooldown time for temperatures below 1~K. The estimated thermal noise is projected to be $\sim 7{\times}10^{-19}$ at 100~mK. However, silicon cavities with crystalline mirror coatings at cryogenic temperatures have shown birefringence correlated frequency fluctuations as well as unknown additional noise mechanisms \cite{yu2023, kedar2023}. We have measured a room-temperature TEM$_{00}$ birefringent mode splitting of about 250 kHz. Understanding and measuring these noise mechanisms will be a key to attaining fractional frequency stabilities beyond state-of-the-art.
\end{abstract}

%
%
%
%
%

\section{Introduction:}
Lasers stabilized to ultrastable optical cavities are required for a wide range of applications such as spectroscopy, gravitational waves detectors, frequency standards and tests of fundamental physics \cite{RevModPhys.87.637}. State of the art optical oscillators based on single crystal silicon cavities exhibit fractional frequency instabilities of few $10^{-17}$ limited by the Brownian thermal noise of the cavity coatings \cite{matei2017}.


Silicon exhibits a very high mechanical quality factor, which increases at low temperatures \cite{Middelmann}. There are two cancellations of the coefficient of thermal expansion (CTE), at $\sim 124$~K and $\sim17$~K \cite{lyon2008}, and below 4~K the  CTE asymptotically approaches zero. Several experiments have shown that fractional frequency instabilities could be attained with such cavities, at 124~K \cite{kessler2012}, 17~K \cite{GILLOT2023}, or below \cite{zhang2017}, along with very low frequency drifts due to the spacer crystalline structure \cite{robinson2019, wiens2020}.

We aim to operate a silicon cavity in a yet unexplored temperature regime below 1~K. We employ crystalline coatings to try and further reduce the cavity thermal noise. Such a cavity will be a unique tool to characterize the cavity materials, including the coefficient of thermal expansion of single-crystal silicon and the behavior of crystalline AlGaAs coatings. It will also be useful to better understand the observed long-term drift of single-crystal silicon cavities, which origin is no yet well understood \cite{wiens2014,robinson2019,wiens2020}. The experimental setup will help to investigate thermal and technical limitations for next generation ultrastable lasers, aiming to realize sub $10^{-17}$ fractional frequency instabilities. This article presents the overall setup we designed for this experiment, along with the projected fundamental and technical noises.

\section{Cryocooler}
Single-crystal silicon cavities have been tested at the 124~K and 17~K turning CTE cancellation points \cite{GILLOT2023} and at temperatures between 1.5~K and 4~K \cite{robinson2019}. The cavity is housed in a dilution cryocooler in order to be able to reach temperatures below 1~K. In addition to the two CTE inversion points documented since the 1970s, a third inversion point has been measured recently close to 3~K \cite{wiens2020}. The ability to control the temperature of the spacer on the 100~mK - 4~K range will therefore be a key to the experiment, and will allow us to test this inversion point as well.

\subsection{Cryocooler setup}

\begin{figure}[h!]
		\center
		\includegraphics[width=10cm]{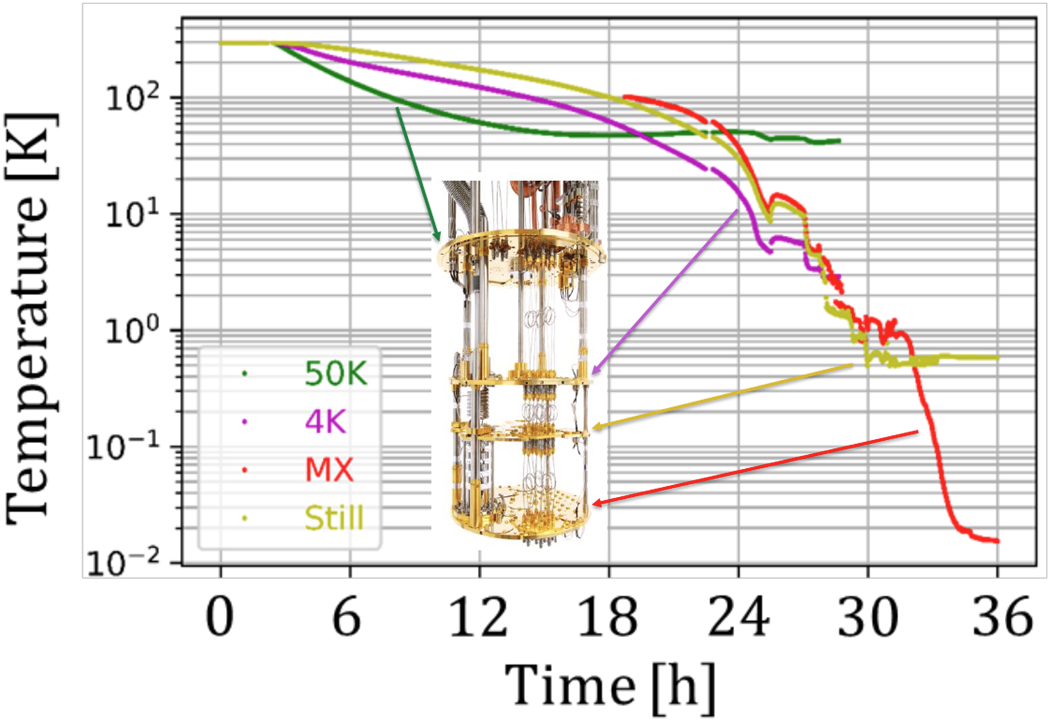}
	\caption{Cooldown of the cryostat. The insert shows the four cryocooler stages. The MX stage reaches the sub-100~mK range in 36 hours.}
	\label{Fig:cooldown}
\end{figure}

Our cryocooler is composed of 4 main stages, shown in Fig. \ref{Fig:cooldown}: the 50~K stage, the 4~K stage, the still stage and the mixing chamber stage (noted MX). The two first stages are based on one closed cycle $^4$He pulse-tube cryocooler. Compression-extension cycle at $\sim 1.4$~Hz of the gas inside ``tubes'' is generated with a compressor unit together with a rotating valve (three positions: open to high pressure, open to low pressure or closed). The first two stages reach temperatures of about 50~K and 3~K.
Then, a Joule-Thomson stage enables temperatures around 1.3~K. To achieve lower temperatures, a dilution of $^3$He in $^4$He is used in a final stage. The plate connected to the mixing chamber can reach a temperature as low as a few mK.

Cylindrical aluminium shields are thermalized on each stage, except for the MX, to significantly suppress radiative heat exchange from the room temperature environment to the cavity. The available space below the mixing plate is a cylinder with a 290~mm diameter and a 300~mm height which sets the maximum dimensions for the cavity spacer. 
Horizontally-oriented optical access is granted through 4 pairs of AR-coated viewports. The window of the vacuum chamber is made of BK7 while thermal shields are equipped with SiO$_{2}$ windows. We estimate that the radiated infra-red power from the 300~K, 50~K and 4~K windows is below 1~µW at the MX stage, well below the available cooling power at 100~mK. 

The cooldown procedure to reach the mK temperature range takes approximately 36 hours without thermal load (cavity) inside and is divided in three main phases (Fig. \ref{Fig:cooldown}). The first one is the pulse-tube $^4$He cryocooling procedure, of roughly 24 hours; the dilution refrigerator can then be started and the $^3$He/$^4$He mixture is first condensed and cools down to about 0.8~K in around 6 hours at the still stage; the $^3$He/$^4$He mixture then separates in two phases and temperatures below 100 mK are reached within roughly 6 hours.

\subsection{Vibration isolation}

Pulse-tube cryocoolers are known to induce vibrations that propagate to the setup through He-circulation hoses and mechanical contacts in the setup. The cavity will be mechanically decoupled from these vibrations through several stages: mechanical grounding of the rotation valve and He hoses through high masses; pneumatic isolation of the cryostat from the cryocooler head; flexible copper braids inside the cryostat that decouple the pulse-tube's vibrations from the science stages.

\subsection{Temperature monitoring}

The temperature of each stage is monitored by at least one sensor (see table \ref{tab:temp_sensors}). The first stage of the pulse tube (50~K) is monitored with a common PT100 sensor and only used in case of cooling issues. Cernox\textregistered \ sensors are used for the 4~K stage and the still plate. They are calibrated for temperatures down to 0.1~K.


Two additional sensors will help monitoring the MX plate and cavity mount temperatures: a calibrated RuO$_{2}$ RX-102B and an uncalibrated RX-202A-CD, with a working range down to 0.01~K. All theses probes are monitored by a multiplexed AC resistance bridge and a temperature controller. To prevent the thermal disturbance caused by the reading of the probe, measurements are performed with a ratio of 3 seconds over 10 seconds.

\begin{table}[h!]
    \centering
        \begin{tabular}{|c|c|c|c|c|}
        \hline
        {Stage} & {type} & {ref.} & {Cal. from} & {to} \\
        \hline
        50~K & PT100 & -- & 310~K & 20~K \\
        4~K & Cernox\textregistered & CX-1010 (LS) & 310~K & 0.1~K \\
        still & Cernox\textregistered & CX-1010 (LS) & 310~K & 0.1~K \\
        MX & RuO$_{2}$ & RX-102B (LS) & 100~K & 0.010~K \\
        MX & RuO$_{2}$ & RX-102B (LS) & 100~K & 0.010~K \\
        Cavity mount & RuO$_{2}$ & RX-202A-CD (LS) & uncal. & uncal. \\
        \hline
    \end{tabular}
    \caption{Temperature sensors installed in the cryostat. LS stands for LakeShore. Cal.: calibration. uncal.: uncalibrated.}
    \label{tab:temp_sensors}
\end{table}


\subsection{Temperature measurements}

We performed preliminary cooling tests for characterizing the cryocooler and temperature probes. The first one, in an unloaded configuration, achieved a temperature of 15~mK, which can be compared to the lowest temperature achieved of 7~mK without windows. During a second run we installed the cavity support and a copper part (see section \ref{sec:copper}) attached to the MX plate where we installed a calibrated RuO$_{2}$ RX-102B and an uncalibrated RX-202A-CD. The measured temperature of this copper block is 20 mK. The temperatures measured by the RuO$_{2}$ sensor on the copper part and the RuO$_{2}$ sensor on the mixing plate agree within 1 mK.

\begin{figure}[h!]
		\center
        \includegraphics[width=12cm]{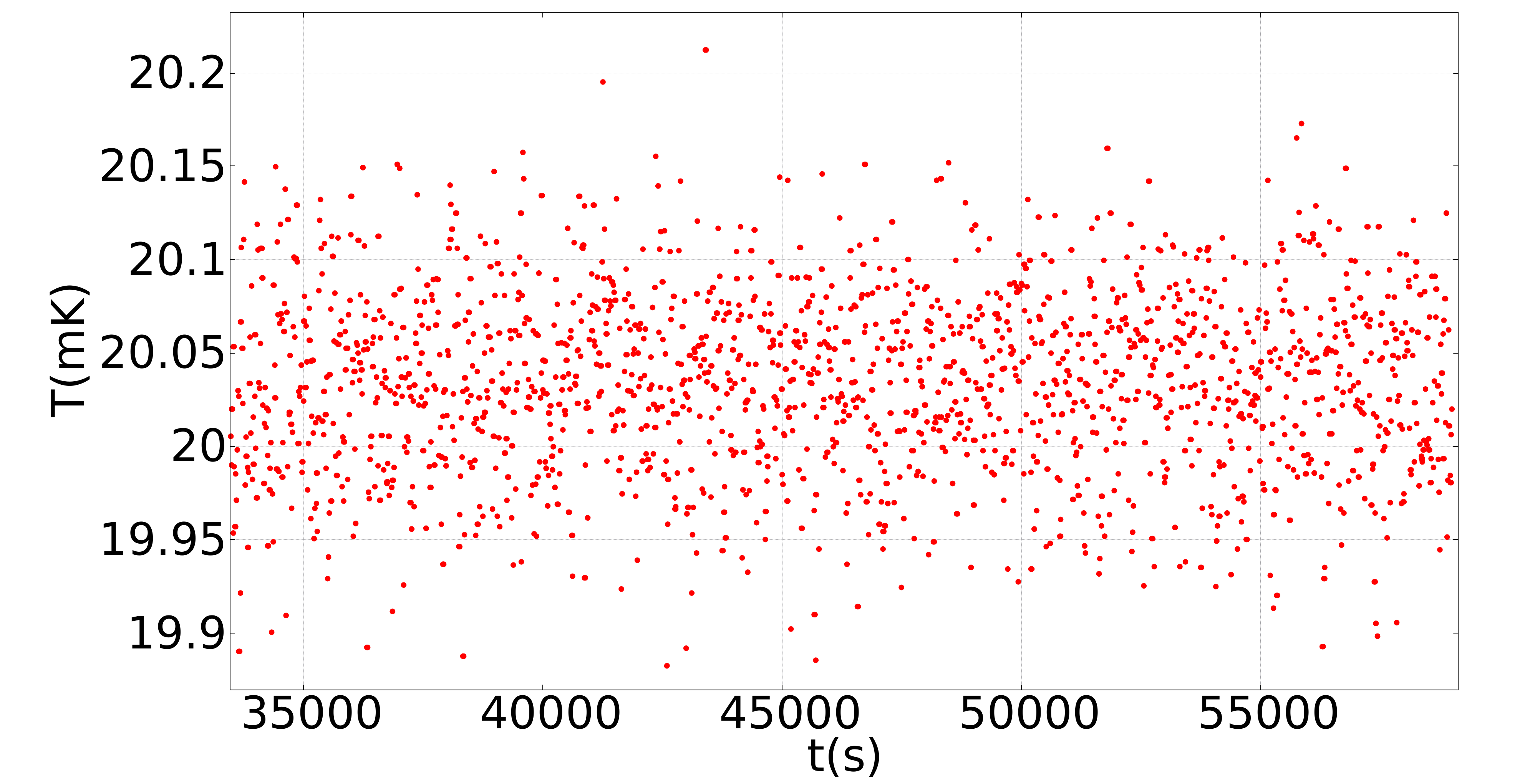}
       \includegraphics[width=12cm]{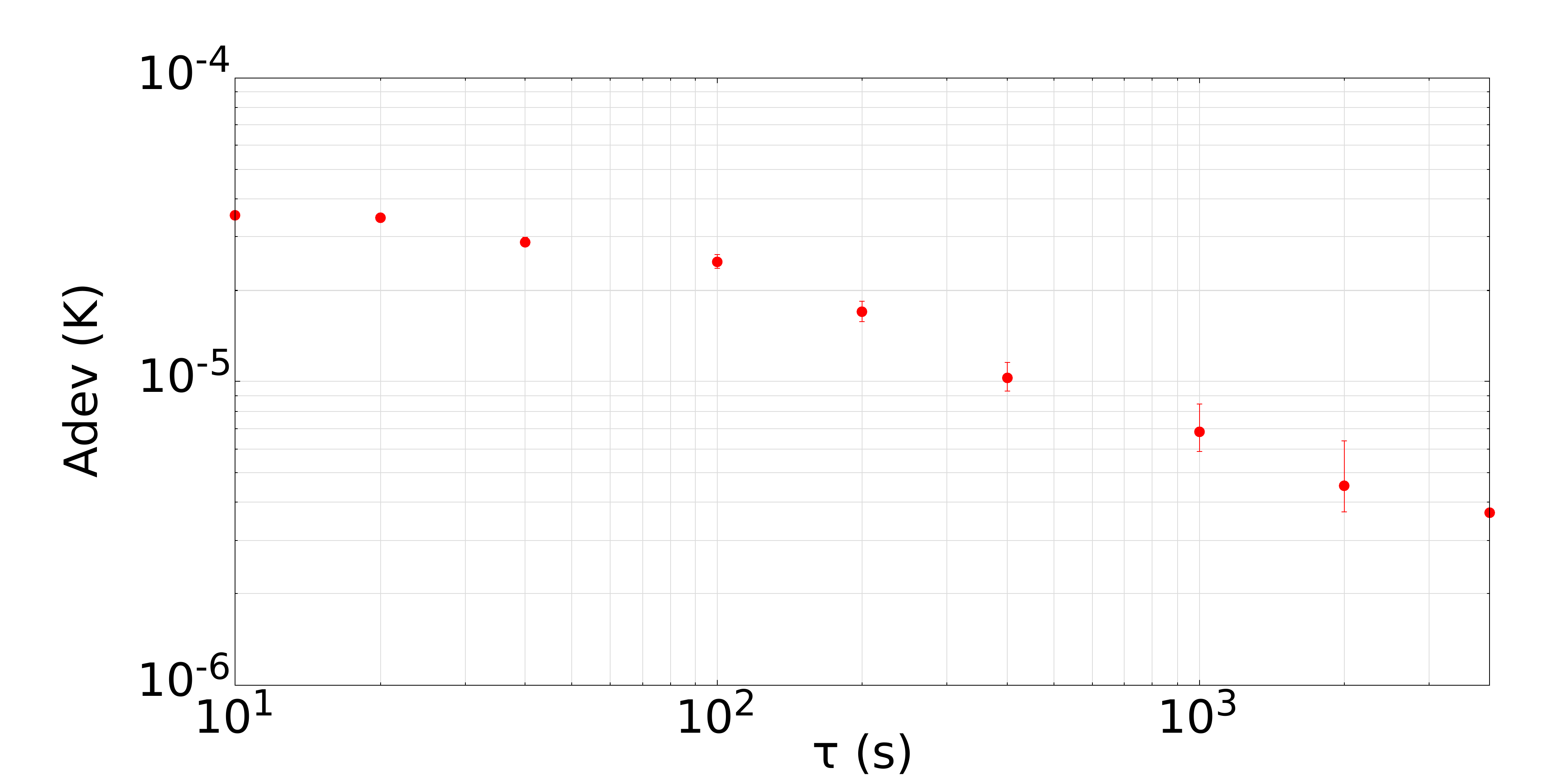}
	\caption{Top: Temperature of the MX stage when temperature stabilization is activated. This is an in-loop measurement. Bottom: Allan deviation of the temperature data.}
	\label{Fig:tempe}
\end{figure}

We tested the temperature control at the MX stage when loaded with the cavity mount. We were able to stabilize the temperature to 20~mK with fluctuations below 0.1~mK, as measured by the in-loop sensor (Fig. \ref{Fig:tempe}).

\section{Cavity}
\subsection{Spacer and mirrors design}
The cavity is composed of a single-crystal silicon spacer and two single-crystal silicon mirror substrates. The dimensions of the cylindrical spacer are 180~mm in length and 200~mm in diameter. We chose the highest possible length given the available space in the MX chamber. The aspect ratio allows us to reduce the radial acceleration sensitivity by cancelling tilts from radial accelerations for this simple cylindrical geometry \cite{jak2009}. The weight of the spacer is about 13 kg. Its large thermal capacity improves the passive filtering of temperature fluctuations.

The optical axes of the spacer and the substrates are aligned to the [111] crystalline axis and include a flat for proper alignment during optical contacting. A custom contacting mask was fabricated to ensure proper centering of the mirrors and proper alignment of the crystalline axis.

The mirrors are equipped with Al$_{0.92}$Ga$_{0.08}$As/GaAs crystalline coatings specified for use at 17~K.  The crystalline coatings are optically contacted to the substrates. Their slow axis is indicated by a flat, which presents a misalignment of less than 6$^{\circ}$ with regards to the substrate's flat. The crystalline orientation of the coatings is most likely similar to those presented in \cite{yu2023}, with the [100] direction of GaAs normal to the mirror surface.


\subsection{Cavity mount}
The cavity mount is made in duralumin in order to reduce its weight. It is attached below the MX plate by four M6 screws to minimize thermal resistance. The cavity is held horizontally on 3 contact points in order to avoid hyper static equilibrium. Further improvement will include simulations of the mount influence on the spacer vibration sensitivity and the search for possible optimal support points such as in \cite{jak2009}.

\begin{figure}[h!]
		\center
		\includegraphics[width=6.5cm]{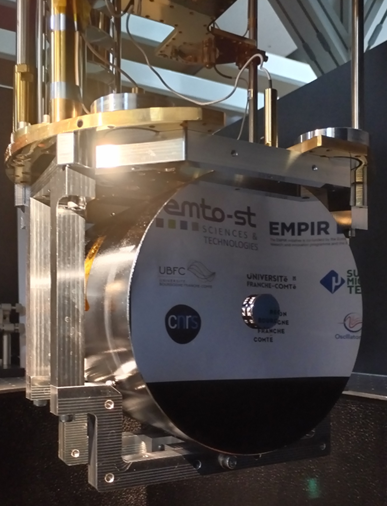}
	\caption{Single-crystal silicon cavity held in its support. The support is directly attached to the MX plate. One of the three supporting spheres is shown on the left bottom corner.}
	\label{Fig:mount}
\end{figure}

Fig. \ref{Fig:mount} shows the cavity held in its support, attached to the MX plate. One of the three supporting spheres made of steel can be seen at the bottom-left. A second one is located in the same plane transverse to the optical axis, while the third one is at the center of the second supporting arch.

\section{Cavity cooling}
\subsection{Cooling by radiation}
As described in the previous section the cavity mount does not enable efficient cooling by heat conduction, as mechanical contacts are voluntarily minimized and the thermal conductance of the 3 contact points is expected to be negligible in the target temperature range. In this configuration, the cavity will be cooled only through the thermal radiation exchange with the surrounding thermal shield. We have performed an analytical estimation of the time required to cool the cavity from $4~$K to $1~$K and from $1~$K to the lowest temperature, corresponding to the Joule-Thomson and dilution refrigeration steps, respectively. For this simple evaluation we assumed: i) the cavity surrounding has reached its thermal equilibrium, ii) the cavity emissivity $\sim 1$, iii) the cavity mass and external surface are $13~$kg and $0.176$~m$^2$ respectively, iv) the silicon heat capacity is given by $C_p=2.5 \times 10^{-4}~T^{3}$~J/kg/K and v) to simplify equation, thermal properties are considered constant within the temperature range. 


In both cases, the cooling time constant is on the order of a few 100 days. We therefore aim to strongly increase the thermal conduction between the cavity and the MX plate.

\subsection{Optimized heat transfer to the cavity} \label{sec:copper}
\begin{figure}[h!]
	\centering
	\includegraphics[width=10cm]{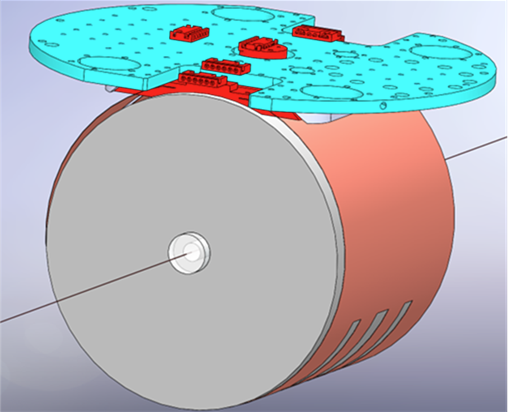}
	\caption{Cavity mount with optimized heat conduction from the cavity to the MX plate: scheme of the copper wrap and its mounting. Red parts are made of solid copper. Three connections are screwed to the MX plate and one is attached to the cavity using a copper foil wrap. Heat conduction is ensured by using soft copper breads (not shown).}
	\label{fig:thermal}
\end{figure}

We have constructed a new cavity mount for optimized heat transfer between the MX plate and the silicon cavity. It is made in copper and will surround the cavity. This thermal support is mechanically decoupled from the setup by using flexible copper braids to thermally contact to the MX plate (Fig. \ref{fig:thermal}).

\section{Preliminary measurements and expected performances}
\subsection{Cavity finesse and birefringence splitting}
\begin{figure}[h!]
	\centering
	\includegraphics[width=12cm]{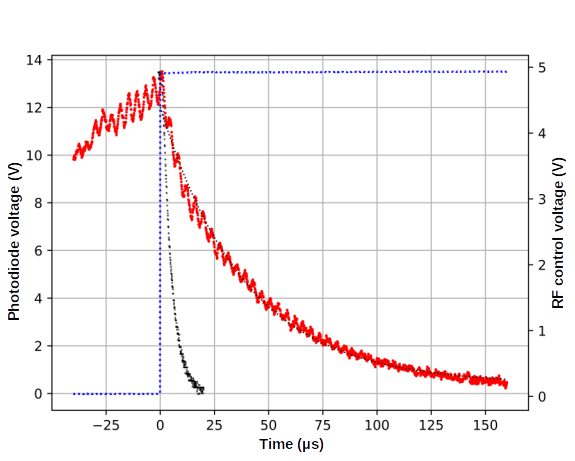}
	\caption{Finesse measurement of the single-crystal silicon cavity in air at room temperature. Red: cavity transmission signal with a superimposed beatnote of the two polarization eigenstates of the cavity. Black: photodiode response without cavity. Blue dotted line: voltage control of the RF switch that turns the light off.}
	\label{fig:finesse}
\end{figure}

The assembled cavity (Fig. \ref{Fig:mount}) shows a finesse of about 220 000 in air at room temperature (Fig. \ref{fig:finesse}) determined by ring-down measurement. The finesse in vacuum and at low temperatures is expected to be higher, as we face residual absorption in air at wavelengths close to 1542~nm and the mirror transmission is optimized for 17~K.

To simultaneously measure the resonance splitting caused by the coatings birefringence, we injected the cavity with linearly polarized light with a projection on both crystalline axes. This results in a beatnote signal superimposed to the typical exponential decay of the ring-down measurement. The TEM$_{00}$ mode splitting due to the birefringence of the coatings is about 250 kHz (Fig. \ref{fig:finesse}), in good agreement to the birefringence measured in other cavities using AlGaAs coatings \cite{yu2023,kedar2023}.


\subsection{Expected performance}
The expected thermal noise of this cavity is listed in table \ref{table:stab}. We use the formula from \cite{numata2004, kessler2012a} for the spacer and substrates combined with the values and formula from \cite{cole2013} for the coatings. 

Assuming peak-to-peak temperature fluctuations of 0.5~mK, the CTE needs to be below $2{\times}10^{-15}$~K$^{-1}$ to reach the $10^{-18}$ level. Several models have been proposed for single-crystal silicon CTE below 1~K. Lyon \textit{et al.} proposed a $4.8\times10^{-13}~T^3$ dependency, where $T$ is the temperature in K \cite{lyon2008}. Wiens \textit{et al.} \cite{wiens2014} proposed a polynomial law derived from experimental values; its accuracy below 1~K is not clear, and it suggests there could be several additional CTE zero-crossings. Those models predict CTE values summarized in table \ref{table:cte}. One can clearly see the advantage of going to the lowest possible temperature. But it is also clear that a better knowledge of the silicon CTE is required in order to accurately determine both, the operating temperature, and the temperature stability constraint. 

\begin{table}[h!]
\centering
\begin{tabular}{|c|c|c|}
\hline
Model & Lyon \textit{et al.} \cite{lyon2008} & Wiens \textit{et al.} \cite{wiens2014} \\
\hline
1~K & $4.8\times10^{-13}$~K$^{-1}$ & $1.7\times10^{-13}$~K$^{-1}$ \\
0.5~K & $6\times10^{-14}$~K$^{-1}$ & $-8.8\times10^{-15}$~K$^{-1}$ \\
0.1~K & $4.8\times10^{-16}$~K$^{-1}$ & $4.0\times10^{-17}$~K$^{-1}$ \\
\hline
\end{tabular}
\caption{Expected coefficient of thermal expansion of single-crystal silicon at cryogenic temperatures, calculated from the model in \cite{lyon2008} and the fit from \cite{wiens2014}.}
\label{table:cte}
\end{table}


Low sensitivity to vibrations will be a key to achieve the targeted performance in a dilution cryostat. We can have a first estimation of these effects assuming a maximum sensitivity of $10^{-10}/$m s$^{-2}$. According to the manufacturers, the typical vibration level in our cryostat can be on the order of 10~nm displacement amplitude; some teams have measured levels of order $5\times10^{-6}$~m s$^{-2}$ at 1~Hz. This would lead to a contribution between $10^{-17}$ and $5\times10^{-16}$ from vibrations. Table \ref{table:stab} shows that such vibrations would limit the performance even at 1~K. To reach the desired levels, the vibration sensitivity and the level of vibrations must be reduced by at least a factor of ten.

It will also be interesting to probe the cavity sensitivity to intra-cavity power fluctuations, which is related to coatings losses and material properties. A sensitivity to optical power was already observed in other experiments \cite{wiens2014, robinson2019, wiens2020, kedar2023, yu2023}.

Two additional contributions have been unveiled recently for cryogenic silicon cavities with crystalline mirror coatings \cite{yu2023, kedar2023}. A birefringence noise, dependent on the intra-cavity optical power and much higher than the predicted Brownian noise, can be averaged out by using the two polarization axis simultaneously. An additional, unknown mechanism dubbed ``residual'' or ``global excess noise'' that is independent of temperature or optical power was observed in two independent experiments \cite{yu2023, kedar2023}, and was measured to be as high as ten times the Brownian noise \cite{yu2023}.

Table \ref{table:stab} summarizes the predicted contributions of thermal noise, global excess noise and thermal fluctuations to the fractional frequency stability at 1~K, 0.5~K and 0.1~K. The ``Global noise'' column accounts for residual excess noise described in \cite{yu2023} substantially different from thermal noise. It remains after averaging the dominating but anti-correlated birefringent noise of the polarization eigenmodes of the cavity and exceeds the expected Brownian thermal noise of the coatings. As this noise shows no strong dependency on temperature, we assumed an identical value of mirror displacement noise as that measured in ref. \cite{yu2023} at 16~K, and scaled it to our cavity length. Measuring the actual value of this global noise at sub-K temperatures will be important, as it would most probably be a major limitation.

\begin{table}[h!]
\centering
\begin{tabular}{|c|c|c|c|c|}
\hline
Temperature & Thermal noise & ``Global noise'' & Temperature fluctuations & Total\\
\hline
1~K & $2.2{\times}10^{-18}$ & $3.6{\times}10^{-17}$ & $2.4\times10^{-16}$ & $2.4\times10^{-16}$ \\
0.5~K & $1.5{\times}10^{-18}$ & $3.6{\times}10^{-17}$ & $3.0\times10^{-17}$ & $4.7\times10^{-17}$ \\
0.1~K & $6.9{\times}10^{-19}$ & $3.6{\times}10^{-17}$ & $2.4\times10^{-19}$ & $3.6{\times}10^{-17}$ \\
\hline
\end{tabular}
\caption{Predicted contributions of thermal noise, global noise and thermal fluctuations to the fractional frequency stability. The total noise is calculated from the quadratic mean of the three contributions.}
\label{table:stab}
\end{table}

\section{Conclusion}
We have presented a dilution cryostat as a testbed to investigate a single-crystal silicon cavity  at sub-K temperatures. In this temperature range the CTE of silicon asymptotically approaches zero and the sensitivity of the cavity to temperature fluctuations decreases significantly. We have optimized the heat transfer to the cavity by copper braids which will accelerate the cooling time in the sub-K temperature range. As a next step we need to mitigate vibrations induced by the pulse-tube cryocooler.

For this cryostat we have setup a 180 mm long silicon cavity employing crystalline AlGaAs mirrors with a projected thermal noise limit in the low $10^{-18}$ range. The cavity room-temperature finesse is 220,000 in air and the cavity exhibits a birefringent mode splitting of 250~kHz. Our setup will allow the study of the recently described birefringent and global noise contributions in AlGaAs coatings even at temperatures below 4~K.

\section*{Acknowledgements}
This work has been supported by the EIPHI Graduate School (contract “ANR-17-EURE-0002”), by the ANR-10-LABX-48-01 First-TF and ANR-11-EQPX-0033 Oscillator-IMP, and by the Région Bourgogne Franche-Comté.

This project (20FUN08 NEXTLASERS) has received funding from the EMPIR programme co-financed by the Participating States and from the European Union's Horizon 2020 research and innovation programme.

\section*{References}
\bibliographystyle{iopart-num}
\bibliography{Biblio}
\end{document}